\documentstyle[12pt]{article}
\makeatletter
\ifcase\@ptsize
 \font\teneufm=eufm10
 \font\seveneufm=eufm7
 \font\fiveeufm=eufm5
 \font\teneusm=eusm10
 \font\seveneusm=eusm7
 \font\fiveeusm=eusm5
\or
 \font\teneufm=eufm10 scaled \magstephalf
 \font\seveneufm=eufm7
 \font\fiveeufm=eufm5
 \font\teneusm=eusm10 scaled \magstephalf
 \font\seveneusm=eusm7
 \font\fiveeusm=eusm5
\or
 \font\teneufm=eufm10 scaled \magstep1
 \font\seveneufm=eufm7
 \font\fiveeufm=eufm5
 \font\teneusm=eusm10 scaled \magstep1
 \font\seveneusm=eusm7
 \font\fiveeusm=eusm5
\fi

\newfam\eufmfam
\newfam\eusmfam
\textfont\eufmfam=\teneufm  \scriptfont\eufmfam=\seveneufm
  \scriptscriptfont\eufmfam=\fiveeufm
\textfont\eusmfam=\teneusm  \scriptfont\eusmfam=\seveneusm
  \scriptscriptfont\eusmfam=\fiveeusm

\def\frak{\ifmmode\let\next\frak@\else
 \def\next{\errmessage{Use \string\frak\space only in math mode}}\fi\next}
\def\frak@#1{{\frak@@{#1}}}
\def\frak@@#1{\fam\eufmfam#1}

\def\sh{\ifmmode\let\next\sh@\else
 \def\next{\errmessage{Use \string\sh\space only in math mode}}\fi\next}
\def\sh@#1{{\sh@@{#1}}}
\def\sh@@#1{\fam\eusmfam#1}

\ifcase\@ptsize
 \font\tenmsa=msam10
 \font\sevenmsa=msam7
 \font\fivemsa=msam5
 \font\tenmsb=msbm10
 \font\sevenmsb=msbm7
 \font\fivemsb=msbm5
\or
 \font\tenmsa=msam10 scaled \magstephalf
 \font\sevenmsa=msam7
 \font\fivemsa=msam5
 \font\tenmsb=msbm10 scaled \magstephalf
 \font\sevenmsb=msbm7
 \font\fivemsb=msbm5
\or
 \font\tenmsa=msam10 scaled \magstep1
 \font\sevenmsa=msam7
 \font\fivemsa=msam5
 \font\tenmsb=msbm10 scaled \magstep1
 \font\sevenmsb=msbm7
 \font\fivemsb=msbm5
\fi

\newfam\msafam
\newfam\msbfam
\textfont\msafam=\tenmsa  \scriptfont\msafam=\sevenmsa
  \scriptscriptfont\msafam=\fivemsa
\textfont\msbfam=\tenmsb  \scriptfont\msbfam=\sevenmsb
  \scriptscriptfont\msbfam=\fivemsb

\def\Bbb{\ifmmode\let\next\Bbb@\else
 \def\next{\errmessage{Use \string\Bbb\space only in math mode}}\fi\next}
\def\Bbb@#1{{\Bbb@@{#1}}}
\def\Bbb@@#1{\fam\msbfam#1}
\def\hexnumber@#1{\ifnum#1<10 \number#1\else
 \ifnum#1=10 A\else\ifnum#1=11 B\else\ifnum#1=12 C\else
 \ifnum#1=13 D\else\ifnum#1=14 E\else\ifnum#1=15 F\fi\fi\fi\fi\fi\fi\fi}
\def\msa@{\hexnumber@\msafam}
\def\msb@{\hexnumber@\msbfam}
\mathchardef\square="0\msa@03

\makeatother
\newcommand{\beq}{\begin{equation}}
\newcommand{\eeq}{\end{equation}}
\newcommand{\ba}{\begin{array}}
\newcommand{\ea}{\end{array}}
\newcommand{\bea}{\begin{eqnarray}}
\newcommand{\eea}{\end{eqnarray}}
\newcommand{\bean}{\begin{eqnarray*}}
\newcommand{\eean}{\end{eqnarray*}}

\newtheorem{theorem}{Theorem}[section]
\newtheorem{prop}[theorem]{Proposition}

\newtheorem{remark}[theorem]{Remark}

\newtheorem{proof}{Proof.}
\newenvironment{pf}{\begin{proof} \rm}{\hfill$\square$ \end{proof}}

\newcommand{\CW}{{\cal W}}

\newcommand{\CL}{{\cal L}}
\newcommand{\CK}{{\cal K}}
\makeatletter
\@addtoreset{equation}{section}

\makeatother

\newcommand{\ZZ}{{\Bbb Z}}

\newcommand{\cmp}[3]{Comm. Math. Phys. {\bf #1} (#2), #3}
\newcommand{\pl}[3]{Phys. Lett. {\bf B #1} (#2), #3}

\newcommand{\lmp}[3]{Lett. Math. Phys. {\bf #1} (#2), #3}

\newcommand{\lanl}[1]{LANL preprint, hep-th/#1}
\newcommand{\jmp}[3]{Jour. Math. Phys. {\bf #1} (#2), #3}
\newcommand{\rref}[1]{(\ref{#1})} 

\newcommand{\del}{{\partial}}

\def\Fdb{{Fa\`a di Bruno}}

\def\dpt#1#2{\frac{\partial #1}{\partial t_{#2}}}
\def\Bdpt#1#2{\displaystyle{\frac{\partial #1}{\partial t_{#2}}}}
\def\H#1{H^{(#1)}}
\def\Hit#1{\Hi^{(#1)}}
\def\Aa#1{A^{(#1)}}
\def\h#1{h^{(#1)}}

\def\a#1{a^{(#1)}}
\def\hi{\tilde{h}}
\def\Hi{\tilde{H}}

\def\pdx{{\partial_x}}

\begin{document}
\begin{titlepage}
\begin{flushright}
Ref. SISSA 131/96/FM
\end{flushright}
\vspace{0.8truecm}
\begin{center}
{\huge Darboux Coverings and Rational\\
Reductions of the KP Hierarchy}
\end{center}
\vspace{0.8truecm}
\begin{center}
{\large
Paolo Casati${}^1$, Gregorio Falqui${}^2$,\\
Franco Magri${}^1$, and
Marco Pedroni${}^3$}\\
\vspace{1.truecm}
${}^1$ Dipartimento di Matematica, Universit\`a di Milano\\
Via C. Saldini 50, I-20133 Milano, Italy\\
E--mail: casati@vmimat.mat.unimi.it,
magri@vmimat.mat.unimi.it\\
${}^2$ SISSA/ISAS, Via Beirut 2/4, I-34014 Trieste, Italy\\
E--mail: falqui@sissa.it\\
${}^3$ Dipartimento di Matematica, Universit\`a di Genova\\
Via Dodecaneso 35, I-16146 Genova, Italy\\
E--mail: pedroni@dima.unige.it
\end{center}
\vspace{0.2truecm}
\abstract{\noindent We use the method of Darboux coverings
to discuss the invariant submanifolds
of the KP equations, presented as conservation
laws in the space of  monic Laurent series in the spectral parameter
(the space of the Hamiltonian densities).
We identify a special class of these submanifolds
with the rational invariant submanifolds entering matrix models of
$2D$--gravity, recently characterized
by Dickey and Krichever. Four examples of the general procedure
are provided.}

\vspace{1.3truecm}
Work supported by the Italian M.U.R.S.T. and by
the G.N.F.M. of the Italian C.N.R
\end{titlepage}
\setcounter{footnote}{0}
\section{Introduction}
Constrained KP
hierarchies were introduced
as symmetry reductions
of the KP hierarchy~\cite{KoStr92,OrSh91,ChLi91},
the symmetry being generated by
suitable combinations of a Baker--Akhiezer
function and an adjoint one.
These integrable systems have been recently
widely studied~\cite{KoStr92,Dik94,Kr95,BoLiXi95,ArNiPa93},
especially in connection with multi--matrix
models of two--dimensional gravity and the theory of
$\CW$--algebras (see, e.g.,~\cite{ArNiPaZi94,BoXi93,Ar95,W-rev}
and references quoted therein).
For instance, their (bi--)Hamiltonian structures and
free--field realizations~\cite{BoLiXi95,ArNiPa93,OeStr93},
and their picture in the Segal--Wilson approach~\cite{Zha94}
are well under control.
Their geometrical structure has been clarified
in~\cite{Dik94,Kr95} by proving that
such hierarchies are the restriction of the KP flows to submanifolds
$\CK_{m,n}$, in which the $n$--th power of the KP Lax operator
$\CL$ factors as the
ratio of two purely differential operators,
$\CL^n= L_{(m)}^{-1} L_{(n+m)}$
(whence the denomination of ``rational reductions'' of the KP theory).

In this Letter we want to present
a different approach to constrained KP hierarchies, based
on the geometrical concept of {\em Darboux covering}. Roughly speaking,
a Darboux covering of a vector field $X$ on a manifold $M$ is
a vector field $Y$ on another manifold $N$ which is doubly related
to $X$ by a pair of maps from $N$ to $M$.
A Darboux covering of the KP equations has been constructed in
\cite{mpz}, by introducing the Darboux--KP (DKP) equations.
There it has been shown that these equations
allow to relate in a natural way
the KP and the modified KP (mKP)
theories, and their discrete (Toda Lattice) counterparts.
Here we want to show how to
use such a covering in the study
of the constrained KP theories, and in the clarification of
the relations among the KP equations, the modified
KP equations\footnote{A possible connection
between constrained KP and mKP hierarchies
was noticed by L. A. Dickey in one
of the preprints of~\cite{Dik94}.}, and the
AKNS--type hierarchies~\cite{Ku95,Kr95,Dik94}.

A peculiarity the reader must be warned of is that our approach to the
KP equations differs from the standard one based on the algebra of
pseudodifferential operators~(see, e.g.,~\cite{DJKM,Dikbook}).
We pursue the
bihamiltonian approach
exposed in~\cite{CFMP2,CFMP3,CFMP4},
in which the basic dynamical variable is the generating function
$h(z)$ of the Hamiltonian densities,
\beq
\label{acca}
h(z)=z+\sum_{i\ge 1} h_i(x)z^{-i},
\eeq
and the KP equations are the associated conservation laws
\beq
\label{conseq}
\del_{t_j} h =\pdx \H{j}.
\eeq
In the same style, the DKP equations are written as a system
\begin{equation} \label{eq1p2200}
\left\{
\begin{array}{lll}
\partial_{t_{j}} h & = & \pdx H^{(j)} \\
\partial_{t_{j}} a & = & a(\Hi^{(j)}-H^{(j)})
\end{array}
\right.
\end{equation}
in a pair of Laurent series $(h,a)$. In our picture, the constrained
KP hierarchies are the restriction of the DKP equations to suitable
invariant submanifolds in the space of pairs $(h,a)$.

The scheme of this Letter is as follows: in
Sections~\ref{sec1} and~\ref{sec2} we recall
from~\cite{mpz} the notion of Darboux covering,
implement it for the KP theory, and find
a family (parametrized by an
integer number)  of invariant submanifolds $S_l$.
In Section~\ref{sec4} we present the Lax map
connecting the standard picture of the KP hierarchy with the one
summarized in equations~\rref{acca} and~\rref{conseq},
and perform the identification
of $S_{n,n+m}=S_n\cap S_{n+m}$ with the Dickey--Krichever rational
submanifolds $\CK_{m+1,n}$.
Finally, in Section~\ref{sec5} we give the simplest examples of
our construction.
The last example,
considering a reduction of the DKP system on the {\em triple}
intersection $S_0\cap S_1\cap S_2$, gives rise to a
non standard one--field reduction of the AKNS system.

\section{Darboux coverings}\label{sec1}
Let us consider the dynamical systems described by the vector fields
$X$, $Y$, and $Z$ on the manifolds $M$, $N$ and $P$, respectively.
We shall say that $Y$ intertwines $X$ and $Z$ if there exists a pair of maps
$\mu:N\rightarrow M$ and $\sigma: N \rightarrow P$ relating $Y$ to $X$
and $Z$, respectively. In the form of a diagram we have
\[\begin{array}{ccccc}
M&\stackrel{\mu}{\longleftarrow}&N&\stackrel{\sigma}{\longrightarrow}&P\\
X&\stackrel{\mu_{\ast}}{\longleftarrow}&Y&\stackrel{\sigma_{\ast}}{
\longrightarrow}&Z
\end{array}\]
If $Z$ coincides with $X$ and $N$ is a fiber bundle over
$M=P$ with canonical projection $\mu$, we shall say that $Y$ is a Darboux
covering of $X$.
So, a Darboux covering is a vector field $Y$ on a fiber bundle over $M$
that intertwines $X$ with itself by means of the canonical projection
$\mu:N\rightarrow M$ and the Darboux map $\sigma:N \rightarrow M$.

The concept of Darboux covering described above may be used to
construct solutions as well as invariant submanifolds of the vector
field $X$.
The construction of solutions  has been detailed in~\cite{mpz}.
In this Letter we will exemplify the use of Darboux coverings
for the reduction of the dynamical system $X$. The reduction process
rests on the elementary remark that any invariant submanifold
${S} \subset N$ of the vector field $Y$ projects into two submanifolds
\begin{eqnarray*}
{S}' & = &   \mu(S) \\
{S}'' & = & \sigma(S)
\mbox{ ,}
\end{eqnarray*}
which are invariant by $X$ on the base space $M$.
Thus, a possible strategy to discover invariant submanifolds of the
vector field $X$ on $M$ is to look for the invariant submanifolds
of its Darboux covering $Y$ on $N$.

\section{Darboux--KP equations}\label{sec2}
Our main example of Darboux covering concerns the KP equations.
Here the manifolds $M$ and $N$, the vector fields
$X$ and $Y$, and the maps
$\mu:N\rightarrow M$ and $\sigma: N \rightarrow P$ are defined as follows.
The manifold $M$ is the infinite dimensional affine space of
monic Laurent series
\beq\label{hdef}
 h(z)   =  z + \sum_{j\ge 1} h_{j}z^{-j},
\eeq
whose coefficients are functions of a space variable $x$.
This space is immersed into the commutative
algebra $L$ of all formal Laurent series
having the form
\beq
k(z)   = \sum_{j\ge -n} k_{j}z^{-j}
\eeq
for some integer $n$.
The manifold $N$ is the product $N=M\times A$, of $M$ with the affine space
$A$ of formal Laurent series of the form:
\beq
\label{adef}
 a(z)  = z + \sum_{j\ge 0} a_{j}z^{-j}.
\eeq
The series in $A$ differ from the ones in  $M$ since they contain
the coefficient $a_0$ as well. This apparently trivial distinction
is basic in the relation between KP and modified KP
equations~\cite{Ku95,mpz}.
The canonical projection
\beq
\mu(h,a) = h  \label{eq1p20}
\eeq
gives rise to the Miura map of the KP theory~\cite{mpz}.

To define the KP equations on $M$ we
introduce the Fa\`{a} di Bruno iterates
$h^{(j)}$ of $h(z)$ defined by
\beq
\label{fdbrr}
\begin{array}{rl}
h^{(0)} & =  1   \\
h^{(j+1)} & =   (\pdx + h) h^{(j)} \mbox{ , }  \forall j \ge 0
\mbox{ .}
\end{array}
\eeq
By linearly combining these iterates in the form
\beq  \label{eq1p14}
H^{(j)} = h^{(j)} + \sum_{l=0}^{j-2} p_{l}^{j}[h] h^{(l)}
\eeq
where the coefficients $p_{l}^{j}[h]$ are allowed to be differential
polynomials in the coefficients $h_j$ of $h$, we select those Laurent
series $\H{j}$
having the asymptotic behaviour
\beq \label{eq1p15}
H^{(j)}=z^j + {\cal O}(z^{-1})\qquad\qquad\mbox{as } z\to\infty.
\eeq
For instance, if we consider
\beq\begin{array}{rl}
\H{2} &=  \h{2}+p_0^{2}[h] \h{0}\\
& = h_x+ h^2+p_0^2[h]\cdot 1\\
& = z^2+ (2 h_1+ p_0^2[h])+ (h_{1x}+ 2 h_2) z^{-1}+ O(z^{-2})
\end{array}\nonumber
\eeq
we immediately see that we have to choose $p_0^2[h]=-2 h_1$ in order
to get rid of the coefficient of $z^0$. Therefore,
\beq
\H{2}=h_x+ h^2 - 2 h_1= z^2 + (h_{1x}+ 2 h_2) z^{-1}+ O(z^{-2}).
\nonumber
\eeq
The Laurent series  $\H{j}$ will be referred
to as the currents of the KP theory.
They define the local conservation laws
\beq
\partial_{t_{j}} h = \del_x H^{(j)} \mbox{ ,}  \label{eq1p18}
\eeq
which are a possible form of the
KP equations.
The well known commutability of the KP flows
follows from the relations
\beq
\partial_{t_j}\H{k}=\del_{t_k}\H{j} \mbox{ ,}
\eeq
proved in~\cite{CFMP4}.\\
To define a Darboux covering of the KP equations we introduce the  map
$\sigma:N\to M$,
\beq
\sigma(h,a) =    h + \frac{a_{x}}{a}  \mbox{ , }  \label{eq1p21}
\eeq
and we consider on $N$ the Darboux--KP equations
(or briefly DKP equations),
\begin{equation} \label{eq1p22}
\left\{
\begin{array}{lll}
\partial_{t_{j}} h & = & \pdx H^{(j)} \\
\partial_{t_{j}} a & = & a(\Hi^{(j)}-H^{(j)})
\end{array}
\right.
\end{equation}
where $\Hi^{(j)}$ is the current $H^{(j)}$ evaluated at the point
$\hi=\sigma(h,a)$.
The proof that the DKP equations provide a Darboux covering of the KP
hierarchy rests on the remark that
if the pair $(h,a)$ is a solution of the DKP
equations, then
$\hi=\sigma(h,a)$ is a solution of
the KP equations since
\beq\label{eqhi}
\del_{t_j}\hi =\del_{t_j}h+\del_x(\Hit{j}-\H{j})=\del_x\Hit{j}.
\eeq
Furthermore, the relation
\beq
\left[\dpt{}{j},\dpt{}{k}\right]\cdot a = a\left(
\dpt{}{j}(\Hi^{(k)}-\H{k})-\dpt{}{k}(\Hi^{(j)}-\H{j})\right)=0
\eeq
shows that
the DKP equations define a commuting hierarchy of vector fields
on $N$.

A less obvious property is that these equations
admit a rich family of invariant submanifolds
$S_{l} \subset N$, defined by the constraints
\beq
 z^{l}a = H^{(l+1)} + \sum_{m=0}^{l} a_{m} H^{(l-m)}
\label{esselle}
\eeq
for any integer $l\ge -1$.\\
In~\cite{mpz} the following two properties
of the DKP equations have been proved.

\begin{prop}\label{prop1}
The DKP vector fields are tangent to the submanifolds $S_l$.
\end{prop}
\begin{prop}\label{prop2}
The modified KP equations are
the restrictions of the DKP equations to $S_0$.
\end{prop}
To explain the last statement, we notice that the submanifold
$S_0$ is defined by the equation
\beq
a(z)=h(z)+ a_0.
\eeq
It shows that $S_0$ is a plane in $N$, which may be parameterized by
$a(z)$. Thus the DKP vector fields  give rise to a system of equations
on the space $A$. In~\cite{mpz} it has been shown that these equations
may be written in the form
\beq
\del_{t_j}a =\del_x\Aa{j}\label{mkp},
\eeq
where the currents $\Aa{j}$ are defined similarly to the currents
$\H{j}$:
one first introduces the Fa\`{a} di Bruno iterates
$a^{(j)}$ by
\beq
\label{fdbrra}
\begin{array}{rl}
a^{(0)} & =  1   \\
a^{(j+1)} & =   (\pdx + a) a^{(j)} \mbox{ , }  \forall j \ge 0
\mbox{ ,}
\end{array}
\eeq
and then one considers the unique linear combinations
\beq
\label{eq1p16}
A^{(j)} = a^{(j)} + \sum_{l=1}^{j-1} q_{l}^{j}[a] a^{(l)}
\eeq
with the asymptotic behavior
\begin{equation}
A^{(j)}=z^j + {\cal O}(z^{0})\qquad\qquad\mbox{as } z\to\infty.
\end{equation}
The resulting equations~\rref{mkp} are one of the possible forms
of the mKP equations~\cite{Ko82,Ku95}.\\
Proposition
\ref{prop1} is our main source of invariant submanifolds of
the KP equations. A simple consequence is:
\begin{prop}   \label{prop3}
The submanifolds
\beq
{S}'_{l,l+n} = \mu({S}_{l}\cap {S}_{l+n})
\eeq
are invariant submanifolds for the KP equations.
\end{prop}
Our task is to show that
they indeed coincide with Krichever's rational
invariant submanifolds.

\section{The Lax map}\label{sec4}
In this section we translate the previous results on the invariant
submanifolds
of the DKP equations in the language of pseudodifferential operators,
with the purpose of comparing our results with those recently suggested
by Dickey and Krichever. The connection is established by means of a
suitable map
\beq
\phi:L\to \Psi D O,
\eeq
relating the commutative algebra
$L$ of scalar--valued Laurent series $h(z)$ to the non--commutative
algebra $\Psi D O$ of pseudodifferential operators on the line.
This map is defined in three steps. First one introduces the
negative iterates
$\h{-1},\h{-2}\ldots$ which are computed by solving backwards
the \Fdb\ recursion relations~\rref{fdbrr}, starting from $\h{0}=1$.
One notices that the full set of the \Fdb\ iterates
$\{\h{j}\}_{j\in\ZZ}$ forms a basis in $L$ attached at the point $h(z)$.
Then one defines the map $\phi$ on the
\Fdb\ iterates $\h{j}$, $j\in\Bbb Z$, by setting
\beq
\label{figen}
\phi(\h{j})=\pdx^j\mbox{.}
\eeq
Finally, one extends the map $\phi$ to the whole space $L$ by
linearity, setting
\beq
\phi(f\h{j})=f\cdot \phi(\h{j})=f\cdot \pdx^j,
\eeq
where $f$ is an arbitrary scalar--valued function independent of $z$.
Actually, $\phi$ is the inverse map to the one introduced in~\cite{Ch78}.
It will be referred to as the {\em Lax map} since it allows
to give the KP theory the usual Lax formulation~\cite{DJKM,Dikbook}.
The Lax operator $\CL$ of the
KP hierarchy is defined as the image
\beq
\CL=\phi(z)
\eeq
of the first element of the standard basis in $L$.
In~\cite{CFMP4} the following property of $\phi$,
\beq
\label{zperh}
\phi(z^k\cdot \h{j})={\del }^j\cdot \CL^k,
\eeq
has been proved. Furthermore, it has been shown that
\beq
\label{popa}
\phi(\H{j})=(\CL^j)_+,
\eeq
where $ (\CL^j)_+ $ denotes the differential part of the $j$--th power of
$\CL$, and that
the KP equations (\ref{eq1p18}) admit the usual Lax representation
\beq
\frac{\del \CL}{\del t_j}=[\CL,(\CL^j)_+].
\eeq
We presently use the map $\phi:L\to \Psi D O$ to write explicitly
the equations of the submanifolds
\beq
{S}'_{l,l+n} = \mu({S}_{l}\cap {S}_{l+n})
\eeq
in the formalism of pseudodifferential operators. We notice
that the submanifold ${S}_{l}\cap {S}_{l+n}$ in $N$ is defined by
the pair of equations
\bea
z^{l}a & = & H^{(l+1)} + \sum_{k=0}^{l} a_{k} H^{(l-k)}\\
z^{l+n}a & = & H^{(l+n+1)} + \sum_{k=0}^{l+n} a_{k} H^{(l+n-k)}.
\eea
By eliminating $a$ we get
\beq
\label{cheqdkp}
 H^{(l+n+1)} + \sum_{k=0}^{l+n} a_{k} H^{(l+n-k)}=
z^n (H^{(l+1)} + \sum_{k=0}^{l} a_{k} H^{(l-k)}).
\eeq
Since the canonical projection $\mu:N\to M$ is simply $\mu(h,a)=h$,
this equation may be seen as the equation defining ${S}'_{l,l+n}$ in $M$.
To get the equation of this submanifold in the algebra
$\Psi DO$, let us use the
map $\phi$ to define the pair of {\em differential} operators:
\bea
L_{(l+1)}:&=&\phi(H^{(l+1)} + \sum_{k=0}^{l} a_{k} H^{(l-k)})\\
L_{(l+n+1)}:&=&\phi(H^{(l+n+1)} + \sum_{k=0}^{l+n} a_{k} H^{(l+n-k)}).
\eea
Then, by using the property~\rref{zperh} we obtain the operator equation
\beq
L_{(l+n+1)}=L_{(l+1)}\cdot \CL^n.
\eeq
The argument above proves
\begin{prop} The image of ${S}'_{l,l+n}$ under $\phi$
is the submanifold of the pseudodifferential operators
$\CL$ verifying the equation
\beq
\CL^n= L_{(l+1)}^{-1}\cdot L_{(l+n+1)}.
\eeq
\end{prop}
In this form, one easily recognizes
that ${S}'_{l,l+n}$ coincides with the rational invariant
submanifold $\CK_{l+1,n}$ introduced by Dickey and Krichever.\\
We end this section by two remarks:\\
1) As in the above-mentioned approach, we recover the
$n$--GD manifolds as a special case of this construction. In fact,
the equation for ${S}_{-1}$ reduces to $a\equiv z$, and hence
the defining relations for ${S}'_{-1,n-1}$ can be solved as
$z^{n}=\H{n},$
which, under the Lax map $\phi$
translates into the well-known operator equation  $\CL^{n}=(\CL^{n})_+.$\\
2) The submanifolds
${S}_{l}\cap {S}_{l+n}$ are merely a special class of invariant
submanifolds of the DKP equations. Other invariant submanifolds
of the KP equations
are obtained by considering intersection of three or more invariant
submanifolds, e.g.,
\beq
{S}^\prime_{l,p,q} = \mu({S}_{l}\cap {S}_{p}\cap {S}_q).
\eeq
Obviously, these multiple intersections can be obtained as
intersections of the double ones, i.e., as intersections of
different Krichever's invariant submanifolds. Nevertheless, in
our formalism they can be more easily handled, as it will be
shown by the example of Subsection~\ref{sec54}.

\section{Some Examples}\label{sec5}
In this section we will give four examples aiming to show how
the formalism of Darboux coverings and Laurent series can be
concretely used to find reductions of the KP equations.
Preliminarily we write explicitly the KP currents $\H{j}$ and the
DKP equations \rref{eq1p22} we shall need henceforth. These equations
play a fundamental role in our construction of the constrained KP
equations. We obtain the latter by reducing, first of all, the DKP
equations on $S_l\cap S_{l+n}$, and then by projecting the reduced
equations on $M$.\par
Since we are interested in the first few equations, we need only
the first three positive Fa\`a di Bruno iterates of $h$:
\beq
\begin{array}{l}
\h{1}=h\\
\h{2}=h_x+h^2\\
\h{3}=h_{xx}+3hh_x+h^3.\\
\end{array}
\eeq
Their linear combinations fulfilling condition
\rref{eq1p15} are the first three KP currents:
\beq
\label{5.1}
\begin{array}{l}
\H{1}=\h{1}\\
\H{2}=\h{2}-2h_1\\
\H{3}=\h{3}-3h_1\h{1}-3(h_2+h_{1,x}).
\end{array}
\eeq
Their expansions in powers of $z$ are:
\beq
\label{5.1b}
\begin{array}{l}
\H{1}=z+h_1z^{-1}+h_2z^{-2}+h_3z^{-3}+O(z^{-4})\\
\H{2}=z^2+(2h_2+h_{1,x})z^{-1}+(2h_3+h_{2,x}+h_1^2)z^{-2}+O(z^{-3})\\
\H{3}=z^3+(3h_3+3h_{2,x}+h_{1,xx})z^{-1}+O(z^{-2}).
\end{array}
\eeq
Then we expand in powers of $z$ the Darboux map $\sigma$. The first
three components are:
\beq
\label{5.1c}
\begin{array}{l}
{\hi}_1=h_1+a_{0,x}\\
{\hi}_2=h_2+(a_1-\frac12a_0^2)_x\\
{\hi}_3=h_3+(a_2-a_0a_1+\frac13a_0^3)_x.\\
\end{array}
\eeq
They allow to compute the currents $\Hit{1}$, $\Hit{2}$, $\Hit{3}$ up
to the order $-1$ in $z$. By inserting into the definition \rref{eq1p22}
of the DKP equations, we finally obtain for the first components:
\par\medskip\noindent
$DKP_2$:
\beq
\label{5.10}
\begin{array}{l}
\Bdpt{h_2}{2}=(2h_3+h_1^2+h_{2,x})_x\\
\Bdpt{h_1}{2}=(2h_2+h_{1,x})_x \\
\Bdpt{a_0}{2}=(2a_1+a_{0,x}-a_0^2)_x\\
\Bdpt{a_1}{2}=(2a_2+a_{1,x})_x+2a_{0,x}(h_1-a_1)\\
\Bdpt{a_2}{2}=(2a_3+a_{2,x}+a_1^2)_x+2a_{0,x}(h_2-a_2)+2a_{1,x}(h_1-a_1)
\end{array}
\eeq
\medskip\noindent
$DKP_3$:
\beq
\label{5.20}
\begin{array}{rl}
\Bdpt{h_1}{3}=&(3h_3+3h_{2,x}+h_{1,xx})_x  \\
\Bdpt{a_0}{3}=&(a_{0,xx}-3a_0a_{0,x}+a_0^3-3a_0a_1+3a_2+3a_{1,x})_x\\
\Bdpt{a_1}{3}=&(3a_3+3a_{2,x}+a_{1,xx})_x
+3a_{0,x}(h_2-a_2)\\ &
+3(a_{1,x}-a_0a_{0,x})(h_1-a_1)+3(a_{0,x}(h_1-a_1))_x
\end{array}
\eeq

\subsection{The AKNS system}\label{sec51}
Let us consider the intersection
\beq
{S}_{0,1} = {S}_{0}\cap {S}_{1},
\eeq
whose equations are
\beq\label{s0s1}
\left\{
\begin{array}{rcl}
a & = & \H{1} + a_0\\
z a & = & \H{2}+ a_0 \H{1} + a_1\ .
\end{array}\right.
\eeq
If we solve the first equation with respect to $h$ and we substitute
into the second one, we obtain the equation
\beq
\label{5.40}
za=\a{2}-a_0\a{1}-(a_1+a_{0,x})
\eeq
for $a(z)$. It shows that $(a_0,a_1)$ are free
parameters on $S_{0,1}$, while all the other coefficients can be
written as differential polynomials in $(a_0,a_1)$. In
particular, we obtain
\beq
\begin{array}{l}
h_1=a_1\\
h_2=a_2=-(a_{1,x}+ a_0 a_1)\\
h_3=a_3=a_{1,xx}+a_{0,x}a_1+2a_0a_{1,x}-a_1^2+a_0^2a_1.
\end{array}
\eeq
Substituting these constraints into the first two DKP equations
\rref{5.10} and \rref{5.20} we obtain
\beq
\label{5.41}
\begin{array}{l}
\Bdpt{a_0}{2}=(2a_1+a_{0,x}-a_0^2)_x\\
\Bdpt{a_1}{2}=-(a_{1,x}+2a_0a_1)_x
\end{array}
\eeq
and
\beq
\label{5.42}
\begin{array}{l}
\Bdpt{a_0}{3}=(a_{0,xx}-3a_0a_{0,x}+a_0^3-6a_0a_1)_x\\
\Bdpt{a_1}{3}=(a_{1,xx}+3a_0a_{1,x}-3a_1^2+3a_0^2a_1)_x.
\end{array}
\eeq
They coincide with the $t_2$ and $t_3$ flows of the $(1|1)$--KdV
theory of~\cite{BoXi93,BoLiXi95},
which gives,
after the identifications
\beq
a_0=-\frac{r_x}{r},\quad a_1=-{r q}.
\eeq
the classical AKNS hierarchy.\\
To see explicitly the connection with the rational reductions,
we simply consider the constraint~\rref{s0s1} as
\beq
\label{cs0s1}
z (\h{1}+ a_0)=(\h{2}+a_0 \h{1}-a_1)
\eeq
following from equation \rref{5.40} of $S_{0,1}$, and we apply
the map $\phi$ of Section~\ref{sec4} to get the equality
\beq
\CL=(\del_x+a_0)^{-1}\cdot (\del^2_x+a_0\del_x-a_1).
\eeq
\subsection{The Yajima--Oikawa system}
The next example concerns the intersection
\beq
{S}_{0,2} = {S}_{0}\cap {S}_{2} ,
\eeq
whose equations are
\beq\label{5.50}
\left\{
\begin{array}{rcl}
a & = & \H{1} + a_0\\
z^2 a & = & \H{3}+ a_0 \H{2} + a_1\H{1}+a_2 \ .
\end{array}\right.
\eeq
They show that the submanifold
${S}_{0,2}$ is parameterized by three fields, say
$(a_0,a_1,a_2)$, and provide the constraints
\beq
\label{5.60}
\begin{array}{l}
h_1=a_1\\
h_2=a_2\\
h_3=a_3=-\frac12(a_{1,xx}+a_{2,x}+2a_0a_{2}+a_1^2).
\end{array}
\eeq
Therefore, the equations of motion of the second flow are
\beq
\label{sfyo}
\left\{
\begin{array}{rcl}
\Bdpt{a_0}{2} & = & \del_x (2 a_1-a_0^2+a_{0,x})\\
\Bdpt{a_1}{2} & = & \del_x (a_{1,x}+ 2 a_2)\\
\Bdpt{a_2}{2} & = & -\del_x (2 a_{2,x} + a_{1,xx}+ a_0(2 a_2+a_{1,x})),
\end{array}\right.
\eeq
as it is easily checked by inserting the constraints \rref{5.60}
into the first DKP equations \rref{5.10}.

The explicit relation with the rational reduction to $\CK_{1,2}$
is easily seen taking into account the expressions \rref{5.1}
for the KP currents $\H{j}$. Indeed we can write the constraint
equations \rref{5.50} as
\beq
\label{chyo}
z^2(\h{1}+a_0)=\h{3}+a_0 \h{2}- 2 a_1\h{1}-(3   a_{1,x}+
2 a_2 +2 a_0 a_1),
\eeq
so that the equations of $\CK_{1,2}$ are
\beq
\label{yaoi}
\CL^2=(\del_x+a_0)^{-1}\cdot(\del_x^3+a_0 \del_x^2- 2 a_1\del_x-(3   a_{1,x}+
2 a_2 +2 a_0 a_1)).
\eeq
Equations~\rref{sfyo} coincide with the equations of the second flow of the
Yajima--Oikawa hierarchy.
The coordinate change relating our picture
to the one of~\cite{KoStr92} can be obtained
comparing~\rref{yaoi} with
\beq
\CL^2= \del_x^2+ u + \psi \del^{-1}\psi^*
\eeq
and it reads:
\beq
a_0=-\frac{\psi_x}{\psi};\quad
a_1=-\frac{u}{2};\quad
a_2=\frac{1}{4}u_x-\frac12\psi \psi^*.
\eeq
We remark that thanks to Proposition~\ref{prop2}, the AKNS
and the Yajima--Oikawa hierarchies can be
obtained as reductions of the {\em modified} KP equations, as well as of
the KP ones.

\subsection{The 2--AKNS system}
We conclude our list of simple intersections
of pairs of submanifolds considering ${S}_{1,2}$, which turns out to
be the 2--AKNS hierarchy, alias a special case of the three--boson
hierarchy described, f.i., in~\cite{DS}.
The defining equations for ${S}_{1,2}$ are
\beq
\label{s1s2}
\left\{
\begin{array}{rcl}
z a & = & \H{2}+ a_0 \H{1} + a_1\\
z^2 a & = & \H{3}+ a_0 \H{2} + a_1\H{1}+a_2\ .
\end{array}\right.
\eeq
Hence, ${S}_{1,2}$, is parameterized by four fields, which we will choose to
be $(a_0,a_1,h_1,h_2)$. Since
\beq
h_3=-h_{1,xx}-2h_{2,x}+h_1^2-a_1h_1-a_0(h_{1,x}+h_2),
\eeq
the $t_2$ flow is given by
\beq
\left\{
\begin{array}{rcl}
\Bdpt{a_0}{2}  & = & \del_x (2 a_1-a_0^2+a_{0,x})\\
\Bdpt{a_1}{2}  & = & 2 a_{2,x}+ a_{1,xx}+2 a_{0,x} (h_1-  a_1) \\
\Bdpt{h_1}{2}  & = &\del_x(2 h_2+h_{1,x})\\
\Bdpt{h_2}{2}  & = & -\del_x (2 h_{1,xx}+3 h_{2,x}-3 h_1^2 +
2 a_0( h_{1,x}+h_2)+2 a_1 h_1)\ .
\end{array}\right.
\eeq
The system~\rref{s1s2} gives, in terms
of the free parameters and of the \Fdb\ polynomials of $h$, the relation
\beq \begin{array}{rcl}
z \cdot(\h{2}+ a_0 \h{1} +(a_1-2 h_1))& = & \h{3}+ a_0 \h{2}+
(a_1-3 h_1)\h{1}+\\
& &+(a_2- 2 a_0 h_1-3 h_2- 3 h_{1,x}).
\end{array}
\eeq
Then the link with the rational reduction to $\CK_{2,1}$ is
obtained as in the previous cases.

\subsection{A triple intersection}\label{sec54}
Finally, we consider the intersection
of the three submanifolds we have so far examined, i.e.,
\beq
S_{0,1,2}:=S_0\cap S_1\cap S_2,
\eeq
which is explicitly given, in terms of \Fdb\ polynomials,
by the system
\beq\label{s0s1s2}\left\{
\begin{array}{rcl}
a & = & h+ a_0\\
z a & = & \h{2}+ a_0 \h{1} +(a_1-2 h_1)\\
z^2 a & = & \h{3}+ a_0 \h{2}+(a_1-3 h_1)\h{1}+(a_2- 2 a_0 h_1-3
h_2- 3 h_{1,x}) \ .
\end{array}\right.
\eeq
We already know (see Subsection 5.1) that $S_0\cap S_1$ is
parameterized by $(a_0,a_1)$. In order to determine $S_0\cap S_1\cap S_2$
we have to impose that $z^2a$ belong to the linear span $H_+$ of
$\{\H{j}\}_{j\ge 0}$ (i.e., to the linear span of
$\{\h{j}\}_{j\ge 0}$). Using the equations of $S_0\cap S_1$ we
obtain
\beq
\label{5.100}
\begin{array}{l}
z^2 a =z (\h{2}+a_0 \h{1}+(a_1-2h_1))\\
\phantom{z^2 a}=z[(\del_x+h)(h)+a_0 h+(a_1-2h_1)]\\
\phantom{z^2 a}=z[(\del_x+h)(a-a_0)+a_0 h+(a_1-2a_1)]\\
\phantom{z^2 a}=(\del_x+h)(za)-z(a_{0,x}+a_1)\ .
\end{array}
\eeq
Since $za\in H_+$ on $S_1$, and $(\del_x+h)(H_+)\subset H_+$ by
the definition of the \Fdb\ polynomials, equation \rref{5.100}
shows that $z^2a\in H_+$ if and only if
\beq
\label{5.110}
a_1=-a_{0,x}.
\eeq
Therefore the restriction of the DKP equations to the invariant
submanifold $S_{0,1,2}$ is a one--field system, which projects
on a one--field reduction of KP. The first flows are immediately
obtained from \rref{5.41} and \rref{5.42}:
\beq
\begin{array}{l}
\Bdpt{a_0}{2}=-(a_{0,x}+a_0^2)_x\\
\Bdpt{a_0}{3}=(a_{0,xx}+3a_0a_{0,x}+a_0^3)_x\ .
\end{array}
\eeq
Since the reduction to $S_{0,1}$ coincides with AKNS (see
Subsection 5.1), we have that the reduction to $S_{0,1,2}$ can
be seen as a reduction of AKNS. In terms of the AKNS variables
$(q,r)$, the constraint \rref{5.110} takes the form
\beq
r_x^2- r r_{xx}+ q r^3 = 0.
\eeq
This is a non--standard constraint of the AKNS hierarchy, which could
be hardly discovered by a direct inspection of the equations.

\subsection*{Acknowledgments}
We thank A. Y. Orlov and J. P. Zubelli
for discussions and suggestions.

\end{document}